\documentclass[12pt,preprint]{aastex}
\usepackage{emulateapj5}
\def\etal{{\it et al.}}
\def\simlt{\lower.5ex\hbox{$\; \buildrel < \over \sim \;$}}

\begin{document}

\title{A revised Cepheid distance to NGC~4258 and a test of the distance scale}
\author{Jeffrey A. Newman\altaffilmark{1}, Laura Ferrarese\altaffilmark{2}, Peter B. Stetson\altaffilmark{3}, Eyal Maoz\altaffilmark{4}, Stephen
E. Zepf\altaffilmark{5}, Marc Davis\altaffilmark{1}, Wendy L. Freedman\altaffilmark{6}, and Barry F. Madore\altaffilmark{6,7}}

\email{jnewman@astro.berkeley.edu, lff@physics.rutgers.edu, Peter.Stetson@hia.nrc.ca, maoz@ism.arc.nasa.gov, zepf@astro.yale.edu, marc@astro.berkeley.edu, wendy@ociw.edu, barry@ipac.caltech.edu}

\altaffiltext{1}{Department of Astronomy, University of California, Berkeley, CA 94720}
\altaffiltext{2}{Rutgers University, New Brunswick, NJ, 08854}
\altaffiltext{3}{Dominion Astrophysical Observatory, 5071 W. Saanich Rd., Victoria, B.C., Canada V8X 4M6}
\altaffiltext{4}{NASA Ames Research Center, MS 245-3, Moffett Field, CA 
94035-1000}
\altaffiltext{5}{Department of Astronomy, P.O. Box 208101, Yale University, New Haven, CT 06520}
\altaffiltext{6}{Observatories of the Carnegie Institution of Washington, 813 Santa Barbara St., Pasadena, CA 91101}
\altaffiltext{7}{NASA/IPAC Extragalactic Database, Infrared Processing and Analysis Center, Jet Propulsion Laboratory, California Institute of Technology, MS 100-22, Pasadena, CA 91125}

\begin{abstract}

In a previous paper (Maoz \etal\ 1999), we reported a Hubble Space
Telescope (HST) Cepheid distance to the galaxy NGC~4258 obtained using
the calibrations and methods then standard for the Key Project on the
Extragalactic Distance Scale.  
Here, we reevaluate the Cepheid distance using the revised Key Project
procedures described in Freedman \etal\ (2001).  These revisions alter
the zero points and slopes of the Cepheid Period-Luminosity (P-L)
relations derived at the Large Magellanic Cloud (LMC), the calibration
of the HST WFPC2 camera, and the treatment of metallicity differences.
We also provide herein full information on the Cepheids described in
Maoz \etal\ 1999. Using the refined Key Project techniques and
calibrations, we determine the distance modulus of NGC~4258 to be
$29.47 \pm 0.09$ mag (unique to this determination) $\pm 0.15$ mag
(systematic uncertainties in Key Project distances), corresponding to
a metric distance of $7.8 \pm 0.3 \pm 0.5$ Mpc and 1.2$\sigma$ from
the maser distance of $7.2 \pm 0.5$ Mpc.  We also test the alternative
Cepheid P-L relations of Feast (1999), which yield more discrepant
results.  Additionally, we place weak limits upon the distance to the
LMC and upon the effect of metallicity in Cepheid distance
determinations.

\end{abstract}

\section{Introduction}

Distances to other galaxies obtained by observations of Cepheid
variables with the Hubble Space Telescope (HST) lie at the core of
most recent efforts to determine the extragalactic distance scale.
The small observed scatter in the relationship between Cepheids'
pulsation periods and luminosities, their large numbers (which allow
many independent measures of the distance to a galaxy), and the
simplicity of the basic physics underlying their variability all have
made them uniquely suitable for this purpose.  As a consequence of
their integral role in establishing the distance scale, however, any
changes in the calibration and application of the Cepheid
period--luminosity (P--L) relationship will affect many other
secondary methods of distance measurement.  Any improvements in
techniques for obtaining Cepheid distances are thus extremely
valuable; but this also means that any changes in this vital link
should be well scrutinized if they are to be adopted.

In a paper describing the final results of the HST Key Project on
the Extragalactic Distance Scale (henceforward, the Key Project), Freedman
\etal\ (2001) make a number of refinements to the techniques used in
earlier papers in the series based upon newly-available information.
First and foremost, microlensing experiments (e.g., Udalski \etal\ 1999) and dedicated efforts (Sebo \etal\ 2001) have greatly
enlarged the set of calibrating LMC Cepheids beyond what had been
studied when the initial Key Project P--L relation was determined (Madore \&
Freedman 1991).  The resulting samples have revealed a modest
correction to the previously adopted P--L slope for $I$ observations;
the $V$--band slope remains unchanged.  In the Key Project
methodology, Cepheid magnitudes are corrected for extinction using
their observed color excess $E(V-I)=(V-I)-(V-I)_0$, where $(V-I)_0$ is
the expected color of an unreddened Cepheid of given period based upon
the LMC--calibrated P--L relations.  This procedure is sensitive to
the reddening, since $A_V=2.45 \times E(V-I)$).  An error in the P--L
slope in $I$ thus propagates into a larger, period--dependent error in
the true distance modulus; in this case, $\Delta \mu_0 = -0.24 (\log_{10}{P}
- 1)$.  Because in more distant galaxies only brighter,
longer-period Cepheids are observable, this generally results in a
distance-dependent revision to HST Cepheid distances which in extreme
cases can reach -0.20 mag.  Further details are given in Freedman et
al. (2001).

Second, our understanding of charge transfer efficiency and related
effects in the WFPC2 CCDs has greatly improved in recent years (e.g.,
Stetson 1998, Dolphin 2000), motivating revisions in the Hill \etal\
(1998) photometric zero points used in most earlier Key Project
papers.  Freedman \etal\ adopt the calibration of Stetson (1998),
which results in a -0.02 mag mean adjustment in $V$ and -0.04 mag in
$I$ from the Hill ``long'' zero points.  Carrying this through the Key
Project procedures for obtaining reddening-corrected distance moduli,
they apply a net correction of -0.07$ \pm 0.07$ mag to distances
obtained using the Hill ``long'' zero point (where the error adopted
reflects a conservative estimate of systematic differences among
recent calibrations).

The third change made in the revised Key Project procedures is the
adoption of Cepheid distance moduli adjusted for metallicity effects
as standard.  The typical metallicity of Cepheids in the LMC with
which the P--L relation is calibrated differs by $\sim 0.5$ dex from
that in many of the fields observed in the course of the Key Project.
Metallicity differences may produce measurable differences in the
colors and magnitudes of Cepheids in those fields from ones found in
the LMC; unfortunately, neither theoretical calculations (as those of
Alibert \etal\ 1999 or Musella 1999) nor observations (cf. Kennicutt
\etal\ 1998, Sasselov et al. 1997, Kochanek 1997, or Nevalainen \&
Roos 1998) have provided a definitive determination of the magnitude,
or even sign, of this effect.  Freedman \etal\ (2001) adopt a
correction to Cepheid magnitudes of $-0.2 \pm 0.2$ mag/dex (based upon
current observational results) as standard.  In previous Key Project
work, distance moduli uncorrected for metallicity effects were
primarily used, though results for a correction of -0.24 mag/dex were
also given.  In their error budget, Freedman \etal\ estimate the
potential systematic error in a typical Key Project Cepheid distance
measurement due to corrections for differences in metallicity from the
LMC to be $\pm 0.08$ mag.

Even with these revisions, important uncertainties remain in the
Cepheid-calibrated distance scale.  Foremost among these is the
distance to the Large Magellanic Cloud (LMC), using which the Key Project
Period--Luminosity relation has been calibrated.  A wide variety of
measurements of the distance to the LMC have been performed in the
past few years.  Many of these studies disagree with each other
statistically, spanning roughly 0.5 mag in distance modulus.
Freedman \etal\ (2001) adopt $18.50 \pm 0.10$ mag as the distance
modulus of the LMC; the uncertainty in the LMC distance modulus leads
directly to a systematic uncertainty in the Cepheid distance scale of
$\pm 0.10$ mag.  After the uncertainties in the LMC distance and
metallicity corrections, the next most significant contribution to the
Key Project systematic error budget is the difficulty in determining
zero points for WFPC2 photometry, which is estimated to lead to a $\pm
0.07$ mag uncertainty in HST Cepheid distance moduli; a variety of
other systematic effects could enter at lower levels.

Given the remaining uncertainties, it is worthwhile to test the
revised Key Project distance scale using a galaxy with a well-known,
primary distance and with a metallicity typical of galaxies observed in the
course of the Key Project.  To be useful, such a test requires that
Cepheids be observed with the same instruments, filters, and parameter
measurement techniques as used for objects in the Key Project sample.
Such a test has already been applied to the original Key Project
distance scale based upon observations of Cepheids in NGC~4258 (Maoz
\etal\ 1999).

The spiral galaxy NGC~4258 (SABbc, $M_B\! =\! -20.0$ mag) presents a unique
opportunity for such a test because of the precision with which its
distance has been measured in a manner independent of the conventional
ladder of astronomical distance scales (Freedman 1998).
Furthermore, its metallicity and distance are similar to those of
typical targets of HST Cepheid programs.  The distance to NGC~4258,
$7.2$ Mpc (corresponding to distance modulus $29.28$ mag), has been
determined using its apparently simple, Keplerian circumnuclear disk
delineated by line-emitting water masers that orbit a supermassive
black hole at its center (Miyoshi \etal\ 1995, Maoz 1995).  
 

This disk was discovered by VLBI observations of water maser emission
from the central region of the galaxy (Miyoshi \etal\ 1995).  It is
about $16$ mas in diameter, $\simlt\!0.1$ mas in thickness, rotates at
speeds of $\approx\!10^3 {~{\rm km}} {~{\rm s}^{-1}}\!$, and is viewed
by us from nearly edge-on.  Most remarkably, the rotation curve of the maser
sources is Keplerian to high precision ($\simlt\!0.5\%$), which
provides very strong evidence for a supermassive black hole at the
galaxy center (Maoz 1995).  The high angular resolution ($0.2~\!$mas)
and spectral resolution ($0.2 {~{\rm km}} {~{\rm s}^{-1}}$) of VLBA
observations (Moran \etal\ 1995) allow a precise definition of the
disk structure and kinematics.  Combining the observed rotation
velocities with the measured centripetal acceleration in the disk
(Greenhill \etal\ 1995) or with the observed proper motions of the maser
sources (Herrnstein \etal\ 1999) permits independent measurements of
the physical size of the disk; comparing these to its observed angular
extent yields the distance to the galaxy via simple geometry.
These distance measurements are subject to only small uncertainties.
The differential systematic error in the maser positions is
$\simlt\!0.04$ mas (Moran \etal\ 1995), translating into a relative distance
error smaller than $0.5\%$. The distance to the masing disk scales
with the disk's inclination as $(\sin i)^{-1}\!$; since the disk is
viewed nearly edge-on, $i\!=\!83^{\circ}\!\pm2^{\circ}$ (Herrnstein \etal\
1996), an error even as high as $4^{\circ}$ would introduce a distance
error of only $1\%$.  Relativistic corrections due to gravitational
redshift and transverse Doppler shift have been taken into account and
are much smaller.  The disk is slightly warped, but the distance
determination is not sensitive to the warp model adopted. The warp
contributes, though, to the small uncertainty in the disk inclination
mentioned above. The total estimated uncertainty in this distance is
$\pm 0.3$ Mpc if the disk is presumed to be circular (which it appears
to be to better than $0.5\%$).  If nonzero eccentricities are allowed,
the uncertainty increases to $\pm 0.5$ Mpc (we adopt this more
conservative value for all further discussions).  The direct, geometric
methods used are believed to have minimal unknown systematic
uncertainties.  The two routes to a distance (proper motions and
accelerations) yield results in agreement with each other to $1\%$.

As described in Maoz \etal\ (1999), we have therefore obtained and
analyzed HST observations of NGC~4258 with the intention of testing,
and potentially of better determining, the zero point of the Cepheid
P-L relation.  In that work, we used the then-standard Key Project
methodologies and calibrations, and found a $\geq1.3\sigma$
discrepancy between the HST Cepheid distance and that obtained from
studies of the masing disk.  In this paper, we perform a similar test
after obtaining a Cepheid distance using the revised Key Project
procedures, allowing an evaluation of the new techniques.  We also
provide herein full information (locations, light curves, finding charts,
etc.) for the Cepheids described in Maoz \etal\ 1999.  We present the
observations, the reduction of the data and the search for Cepheids in
$\S 2$, the resulting derivation of a Cepheid distance in $\S 3$, and
the implications in $\S 4$.

\section{Observations, Data Reduction, and Searches for Variable Stars}

We observed a portion of NGC~4258 using the Wide Field and Planetary
Camera 2 (WFPC2) instrument and the Hubble Space Telescope on 11
epochs in 1998.  The data were acquired with an optimal power-law
spacing between them as described in Freedman \etal\ (1994) and Madore
\& Freedman (2001).  The $F555W$ and $F814W$ filters were used for
a combined total of one orbit at every epoch, with two frames obtained
using each filter to limit the effects of cosmic rays; the exposure
time per frame was 500 seconds.  A journal of observations is given in
Table 1.  To simplify analysis, a fixed orientation was maintained for all
epochs.  The field observed is superimposed on a ground-based image of NGC~4258 in Fig. 1.

\subsection{Photometric Reductions}

We have obtained photometry from these data with two commonly-used
software packages: DAOPHOT/ALLFRAME and DoPHOT.  For both analyses,
the data were first preprocessed via the standard Space Telescope
Science Institute pipeline (Holtzman \etal\ 1995).  For the
ALLFRAME photometry each frame was also corrected for vignetting and
geometrical effects on the effective pixel area as described in
Stetson \etal\ (1998).  Photometry was then performed on each of the
data frames using the DAOPHOT II/ALLFRAME package (Stetson 1987).
ALLFRAME fits a predefined point-spread function (PSF) to all stars on
a frame and iteratively determines their magnitudes.
The procedures used throughout were similar to those of the Key
Project (see Stetson \etal\ 1998).  The photometric zero points of
Stetson (1998) were used.  Aperture corrections for each frame were
derived using a set of bright, isolated stars.

In addition to the Cepheid analysis described below, we also attempted
to determine a tip of the red giant branch (TRGB) distance (Lee \etal\ 1993) to NGC~4258
using the ALLFRAME photometry for WF2 and WF3.  However, our
observations were insufficiently deep and contamination by other
stellar populations too great to allow any convincing detection of the
TRGB.


The DoPHOT photometry was performed using a variant of the DoPHOT
package (Schechter \etal\ 1993, Saha \etal\ 1994) which was developed
especially to deal with the photometry of undersampled images such as
those obtained with the HST.  In addition to the data processing
mentioned earlier, pairs of cosmic ray split images were combined prior
to performing the photometric reduction with DoPHOT. The algorithm
used for this is designed to reject cosmic ray events; particular
care is taken to ensure that the photometry of real objects is
preserved.  Further discussion of the application of DoPHOT to
photometry of HST images can be found in Saha \etal\ (1996a),
Ferrarese \etal\ (1996, 1998), and Hill \etal\ (1998).

The photometric calibration adopted again follows Stetson (1998), as
per the revised Key Project procedures of Freedman \etal\ (2001).  The
limited number of bright, uncrowded stars prevented us from deriving
reliable spatially-dependent aperture corrections for DoPHOT.  We
therefore adopted corrections obtained from observations of an
uncrowded field in the Leo I dwarf galaxy.  Aperture photometry
conducted on the NGC~4258 individual frames, as well as on a deep
frame obtained by combining all available epochs, produces results in
agreement with the Leo I aperture corrections at the 0.03 mag level,
which we therefore adopt as a measure of the uncertainty in the
correction themselves.

\subsection{Variable Star Searches}

Searches for Cepheid variables using the ALLFRAME photometric
measurements were conducted using two different algorithms.  One of
these selects candidate variables via a modified Welch-Stetson test
and performs a nonlinear fit of template Cepheid light curves to their
photometry to assess their variability and determine their parameters;
see Stetson (1996) and Stetson \etal\ (1998) for further description.  The
other linearly fits template light curves defined on a grid in period
and phase to all stars with well-determined photometry; Cepheids are
then identified by a set of criteria that are effective at eliminating
nonvariables.  This algorithm is described in more detail in Newman et
al. (1999).  Both searches independently yielded similar sets of
candidate Cepheid variables and parameter estimates in good agreement.
These algorithms in combination yielded 21 potential Cepheid
variables.


A search for variable stars was also conducted using DoPHOT magnitude
measurements for the $V$ and $I-$band frames independently following
the procedure described by Saha \& Hoessel (1990). We required that a
star be detected at at least 8 of the 11 epochs in order to be checked
for variability. We also excluded all stars in crowded regions by
rejecting candidates that had a companion contributing more than 50\%
of the total light within a two-pixel radius. A detailed discussion of
the search procedure can be found in Ferrarese \etal\ (1996).  A star
meeting the above constraints was flagged as a variable if $\chi^2_r
\ge 8$ or $\Lambda \ge 3$ where $\chi^2_r$ and $\Lambda$ are as used
in Saha \& Hoessel (1990).

Several spurious variables were detected by this procedure as a
consequence of non-Gaussian sources of error and various anomalies in
the images (e.g., residual cosmic ray events) along with the crowding
referred to earlier. Each variable star candidate was visually
inspected by blinking several of the individual frames against each
other. The best period for each variable was selected by phasing the
data for all periods between 3 and 60 days in incremental steps of 0.1
days.  Although in most cases the adopted period corresponds to
a minimum value of the phase dispersion, in a few cases an obvious
improvement of the light curve was obtained for a slightly different
period.  The DoPHOT analysis identified a total of 28 potential
Cepheids.

\subsection{DoPHOT-ALLFRAME Comparison}

All Cepheids found in either dataset were on Wide Field (WF) chips 2 or 3;
we therefore will restrict our discussion to these for the remainder
of this paper.  On WF2, the agreement of DoPHOT and ALLFRAME results
was well within the expected errors in the aperture corrections used;
the mean difference between ALLFRAME and DoPHOT magnitudes for 24
bright, isolated stars was $0.026 \pm 0.049$ (standard deviation;
standard error of the mean 0.009) mag for $V$, and $0.015 \pm
0.049$ mag for $I$.  For WF3, the mean difference for 30 stars was
$0.025 \pm 0.027$ mag for $V$, and $0.088 \pm 0.046$ mag for $I$.
Mean magnitudes for Cepheids yielded results consistent with these to
within 1$\sigma$, albeit with much larger standard errors due to their
fainter magnitudes.  

We believe that the WF3/$I$ results are an aberration closely related
to the discrepant distance moduli obtained from ALLFRAME magnitudes on
the two chips (q.v. $\S$ 3).  This anomoly is plausibly accounted for
by the difficulties of determining aperture corrections in the
observed fields, which contain few bright, isolated stars.  If we
presume that this is the case, the WF3/$I$ ALLFRAME photometry may be
corrected by bringing it onto the same system as the WF2 and WF3/$V$
ALLFRAME magnitudes; i.e., adjusting the WF3/$I$ ALLFRAME magnitudes
to be $0.022 \pm 0.0035$ mag fainter than DoPHOT (the average
difference from the other three chip/filter combinations), rather than
$0.088 \pm 0.008$ mag without any correction.  Averaging the
ALLFRAME-DoPHOT differences from those three cases may be justified by
the fact that errors in aperture corrections, etc. are generally
highly correlated between $V$ and $I$ and from one WF chip to another;
and indeed, the ALLFRAME-DoPHOT offsets for WF2/$V$ and $I$ and
WF3/$V$ are all quite consistent, agreeing to within $0.01$ mag.
Therefore, in addition to presenting results for unmodified ALLFRAME
photometry, in the next section we also provide distance measurements
obtained from ALLFRAME data ``corrected'' by subtracting $0.066 \pm
0.009$ mag from WF3/$I$ magnitudes.

\section{The Cepheid P-L Relations}

We have identified and determined light curves, periods, mean
magnitudes, and colors for 15 definitive Cepheids in NGC~4258.  All of
these stars fulfill four criteria: they are identified as variable by
all three search techniques, they fit a Cepheid template light curve with
reasonable $\chi^2$, they visibly vary in blink comparisons in both $F555W$
and $F814W$ images, and they have negligible statistical probability of
being misidentified nonvariables.  

A variety of methods exist for determining the mean magnitudes of
Cepheids.  In this paper, we use the intensity-weighted mean magnitude
obtained from a fit to the Cepheid light curve using the templates of
Stetson (1996), also known as ``template fit'' mean magnitudes.  This
is the preferred method for the Key Project, providing a robust method
of determining mean magnitudes analogous to those obtained for more
densely sampled datasets (such as the LMC Cepheids used for
calibrating the P--L relation).  We note that adopting other standard
magnitude averaging methods changes the NGC~4258 distance modulus
obtained by no more than 0.04 mag.  We use the Cepheid periods
determined in the DoPHOT analysis for all distance modulus
calculations, as these yielded smaller scatter about the P--L relation
for all datasets and averaging methods than alternatives (e.g.,
periods determined from template fits), likely reflecting the fact
that those periods were refined by hand when necessary to improve the
Cepheid light curves, rather than being determined solely by an
automated algorithm.  We have adopted the DoPHOT photometry for all
major conclusions reported here, since as discussed in $\S$ 3,
ALLFRAME photometry yielded internally discrepant distances from
Cepheids on the two WFPC2 chips used; however, as an additional
consistency check we also provide ALLFRAME values in much of what
follows.  We present the locations and characteristics of the Cepheids
found in Table 2, and complete DoPHOT photometry for those stars in
Table 3.  The positions of the Cepheids are shown in Fig. 2, and
detailed finding charts may be found in Fig. 3.  DoPHOT light curves
for the Cepheids found are depicted in Fig. 4.

In accordance with the revised Key Project methodology, we adopt the
P-L relation slopes of Freedman \etal\ (2001) and only fit for
differences in the zero point.  Their LMC calibration yields mean
absolute magnitudes for Cepheids
\begin{small}
\begin{eqnarray}
\bar{M_V} & = & -2.760(\pm0.03) (\log_{10}{P}-1)-4.218(\pm 0.02) \\
\nonumber\bar{M_I} & = & -2.962(\pm0.02)(\log_{10}{P}-1) -4.904(\pm
0.01) ,\\ \nonumber\end{eqnarray}\end{small} where $\bar{M_V}$ and $\bar{M_I}$
are the intensity-weighted mean absolute Johnson $V$ and Kron-Cousins
$I$ magnitudes of the star and $P$ is its period in days.  The same relations have been used by Freedman \etal\ (2001) and Macri \etal\ (2001).  Fitting the
observed magnitudes of Cepheids in NGC~4258 with such relations yields
a measurement of the apparent distance modulus of the galaxy.  We have
done this fitting with the standard Key Project processor, which
determines overall distance moduli as an unweighted mean of the values
for individual stars.  From the difference between the absolute
magnitudes of LMC Cepheids (for an assumed LMC distance modulus of
18.50 mag) and the observed magnitudes of NGC~4258 Cepheids we then
may derive a $V$ or $I$ distance modulus to NGC~4258.  We present the
resulting distance moduli for both ALLFRAME and DoPHOT photometry in
Table 4, both for the entire sample and the subsets of Cepheids on
either Chip 2 or Chip 3 (which can indicate the presence of
catastrophic photometric or other errors).  The DoPHOT NGC~4258 P-L relations are plotted in Figure 5, and those from ALLFRAME in Figure 6.

In Key Project procedures, the colors of the observed Cepheids
(compared to those of a calibration set of such stars in the LMC whose
reddening has been assumed) are then used to correct for line-of-sight
of extinction assuming a Cardelli \etal\ (1989) reddening law.  This
may be done by comparing the distance moduli obtained in the $V$ and
$I$ bands; the difference of the two measures the average value of
$E(V-I)$.  Because the Key Project processor uses only unweighted
averages with no rejection, this is equivalent both to applying the
extinction correction star-by-star and then averaging and to fitting
the P-L relation for a reddening-corrected ``Wesenheit'' magnitude,
$\bar{W}=\bar{V}-2.45(\bar{V}-\bar{I})$ (Madore 1982, Madore \&
Freedman 1991).  This method yields $E(V-I)=0.20 \pm 0.04$ mag and an
extinction-corrected distance modulus of $\mu_0=29.40 \pm 0.06$ mag
from DoPHOT photometry for all Cepheids; for the ALLFRAME photometry,
the corresponding numbers are $0.19\pm 0.04$ and $29.53 \pm 0.07$
mag.  The true moduli for the subset of Cepheids on either WF2 or WF3 are also listed in Table 4.
Note that in the ALLFRAME dataset, Cepheids on
the two chips yield extinction-corrected distance moduli differing by
0.25 mag, or 1.9$\sigma$; the DoPHOT moduli differ by only 0.07 mag
(0.5$\sigma$).  Furthermore, the value for WF2 is quite consistent
with that obtained from DoPHOT photometry, both for single chips'
samples and overall; this makes the WF3 ALLFRAME results particularly suspect.

The bulk of this discrepancy is eliminated if we perform the WF3/$I$
correction described in $\S$ 2.3.  That adjustment of WF3/$I$ magnitudes
by $0.066 \pm 0.009$ mag reduces the overall distance modulus yielded by
Cepheids on WF3 by $0.162 \pm 0.021$ mag.  Such a correction reduces
the ALLFRAME WF2/WF3 discrepancy to 0.09 mag, a 0.7$\sigma$
difference.  As seen in Table 4, this correction would leave
ALLFRAME $V$ distance moduli unchanged, alter the WF3 and overall
average $I$ moduli to 29.77 mag, and reduce the extinction-corrected
distance modulus to $29.48 \pm 0.09$ mag for WF3, or $29.44 \pm 0.065$
mag overall (where the errors include the propagated error from the
uncertainty in the ALLFRAME-DoPHOT differences, 0.021 mag, added in
quadrature).  These values are in much better agreement with those
obtained from DoPHOT photometry (for which the overall,
extinction-corrected distance modulus was $\mu_0=29.40 \pm 0.06$ mag).
Applying this correction also reduces the scatter in the overall,
extinction-corrected ALLFRAME distance modulus from 0.35 mag to 0.25
mag.

Freedman \etal\ (2001) find that differences in metallicity have an
effect on extinction-corrected Cepheid distances of 0.2$\pm 0.2$
mag/dex.  Using the fits to data on NGC~4258 HII regions from
Zaritsky, Kennicutt, and Huchra (1994), we estimate the metallicity in
our HST fields to be 12+log(O/H)=$8.85 \pm 0.06$, 0.35 dex higher than
that adopted for Cepheids in the LMC.  This leads to a correction of
+0.07$\pm 0.07$ mag to the distance moduli we have derived.  The
revised Key Project procedures adopt metallicity-corrected values for
the distance modulus as their primary estimate (in contrast to
previous practice); we thus do likewise, and obtain a final true
modulus to NGC~4258 of 29.42 mag.  Because the metallicity of NGC~4258
is quite typical of Key Project targets, we treat the uncertainty
in the metallicity correction as a systematic uncertainty in the
distance scale.

\subsection{Uncertainties}

We present an error budget for our measurement of the distance to
NGC~4258 in Table 5.  In addition to the random errors determined in
the course of fitting P--L relations, uncertainties in the aperture
corrections used are also random between different Cepheid target
galaxies.  From studies of globular clusters and Key Project galaxies,
we expect these to be approximately 0.05 magnitude or less in both $V$
and $I$, and highly correlated between the two bands; the combined
uncertainty due to the aperture corrections in the reddening-corrected
distance modulus would then still be 0.05 mag.  Differences between
ALLFRAME and DoPHOT photometry of bright stars are much smaller than
this in all cases save WF3/$I$.  Since it appears that the difference
between the overall DoPHOT and uncorrected ALLFRAME distance primarily
reflects a correctable error in the WF3/$I$ ALLFRAME photometry alone,
we estimate that photometric errors in the DoPHOT results which are
unique to our study of NGC~4258 may constitute 0.05 mag. Adding all
potential sources in quadrature, the total random uncertainty in our
determination of a Cepheid distance to NGC~4258 ($R_{tot}$ in Table
5) is 0.09 magnitude.

This measurement is also subject to a number of potential sources of
systematic error that affect Key Project distance determinations
similarly, as described in Table 5; their possible contributions
have been estimated to total $\pm 0.15$ mag (Ferrarese \etal\ 1999,
Freedman \etal\ 2001). For those potential systematic errors which
affect all Cepheid distances obtained in the same manner as ours
uniformly, we have adopted the uncertainty estimates of the Key
Project (Freedman \etal\ 2001); more detailed descriptions may be found
therein.

We thus obtain a Cepheid distance modulus to NGC~4258 of $29.47
\pm 0.09$ mag (unique to this determination) $\pm 0.15$ mag
(systematic uncertainties in Key Project distances), corresponding to
a metric distance of 7.8 $\pm 0.3$ Mpc $\pm 0.5$ Mpc.  When treated
in the same way, the uncorrected ALLFRAME results yield a distance
modulus of $29.60 \pm 0.10$ mag, corresponding to a metric distance
of $8.3 \pm 0.4$ Mpc, while the corrected ALLFRAME results yield a
distance modulus of $29.51 \pm 0.09$ mag, corresponding to a metric
distance of $8.0 \pm 0.3$ Mpc.  The distance to NGC~4258 derived
from observations of Cepheids using the revised Key Project
methodologies is thus not significantly greater than the maser
distance of 7.2 $\pm ~0.5$ Mpc (Herrnstein \etal\ 1999).

\section{Implications}

In assessing the validity of the calibration of the Cepheid distance
scale using NGC~4258, we must consider how significant the difference
we have found is.  First of all, we may examine whether the
differences are consistent within the random and systematic error
budgets of the two processes; i.e., whether previously considered
sources of error are sufficient to account for what we have found.
The Cepheid and maser distances differ by 0.9$\sigma$ if we add in
quadrature our measurement uncertainty of 0.3 Mpc, the Key Project
systematic error estimate of 0.5 Mpc, and the maser distance error
estimate of 0.5 Mpc; potential systematic errors in either technique
do not appear to have been underestimated.  In performing a test of
the validity of the revised Key Project distance scale, however, we
must consider what sort of a discrepancy we can measure.  All HST
Cepheid distances following the revised techniques will share the
systematic errors that affect our results.  Thus, we should consider
only the random errors unique to this measurement and those errors
affecting the maser distance in determining whether a recalibration
would be an improvement.  In that case, we find a 1.2$\sigma$
difference, as opposed to a 1.6$\sigma$ difference if the results of
Maoz \etal\ (1999) were evaluated with the same error budget.  Put
differently, if we were to presume the maser distance to NGC~4258 is
correct and recalibrate the distance scale based upon our Cepheid
observations of NGC~4258, all revised Key Project distances would have
to be reduced by 0.19 mag, increasing the resulting measurements of
the Hubble constant by $10\%$; the total systematic error budget for the
new calibration would be 0.16 mag (8\%), slightly greater than
that resulting from the current, LMC-based methods.  Modest changes in
calibration or methodology would be sufficient to bring the Cepheid
and maser distances into substantially better agreement; for instance,
the LMC distance need only be reduced by a few hundredths of a magnitude, well within the Key Project estimate of its uncertainty, to bring the difference below 1$\sigma$.

We can similarly use our measurement of the distance to NGC~4258 to
test other calibrations of the Cepheid P-L relation.  In particular,
we have applied the Milky Way-based calibration of Feast (1999), which corresponds to:
\begin{small}
\begin{eqnarray}
\bar{M_V} & = & -2.81(\pm0.06) (\log_{10}{P}-1) - 4.26(\pm 0.05) \\
\nonumber\bar{M_I} & = & -3.07 (\log_{10}{P}-1) - 4.89,\\
\nonumber\end{eqnarray}\end{small} following the conventions of equation 1.
Note that the slopes adopted in deriving these relations are
substantially different from those used by Freedman \etal\ (2001).
Although the Cepheids in Feast's calibration are not at uniform
distance, making their use to calibrate the P-L relation somewhat more
complicated, they are independent of any assumptions about the LMC
distance and of very similar metallicity to the fields we have
studied.  Using these relations leads to a Cepheid distance modulus
for NGC~4258 of $29.51 \pm 0.06$ (random) mag, 0.23 mag greater than
the maser result; a metallicity correction of -0.2 mag/dex increases
this by only 0.01 mag.  This is a total discrepancy of 1.3$\sigma$ if we
consider all possible errors and assume a systematic uncertainty of
0.05 mag in Feast's calibration, 1.2$\sigma$ for 0.1 mag, or 1$\sigma$ for
0.15 mag uncertainty.  If, as above, we compare the discrepancy to
what we can measure, we find that the maser and Cepheid distances are
1.5$\sigma$ apart; the calibration of Feast (1999) might be improved
by rereferencing it to NGC~4258.

Some authors (e.g. Gibson 2000) have used the results of Maoz et
al. (1999) to estimate the distance to the Large Magellanic Cloud,
based upon the assumption that the maser distance to NGC~4258 is
correct and that any differences between the Cepheid distance to that
galaxy and the maser distance are due to an error in the distance to
the LMC assumed in calibrating P--L relations.  Such a procedure is
subject to many sources of uncertainty; the random uncertainty in the
resulting LMC distance will correspond to the sum in quadrature of the
random uncertainties in the maser and Cepheid distances to NGC~4258.
Furthermore, the uncertainty due to potential systematic errors in
that procedure includes contributions from all possible systematics in
the maser and Key Project error budgets, save the distance to the LMC
itself.  If we nevertheless proceed in this manner, we determine an
LMC distance modulus of 18.31 $\pm$ 0.11 (random) $\pm$ 0.17
(systematic).

Similarly, we may use the maser and Cepheid distances to NGC~4258 to
place limits on the effect of metallicity on Cepheid luminosities.  In
that case, we should consider all random and systematic uncertainties
in the maser or Cepheid distances save those due to metallicity
effects; they total 0.21 mag.  Thus, the Cepheid distance becomes
1$\sigma$ discrepant from the maser distance if there is a total
metallicity correction of +0.09 mag to the uncorrected Cepheid
distance, and 2$\sigma$ discrepant with a correction of +0.30 mag.  As
the metallicity of NGC~4258 is $0.35 \pm 0.06$ dex greater than that
of the LMC, we thus can limit the effect of metallicity on Cepheid
mean magnitudes to -0.26 mag/dex at 1$\sigma$, or -0.9 mag/dex at
2$\sigma$.  

NGC~4258 has provided the most stringent geometrical test of the
revised Key Project distance scale so far.  The 0.26 mag discrepancy
between the maser distance and the Cepheid distance to NGC~4258
obtained with the original Key Project methods diminishes to 0.19 mag
when the revisions of Freedman \etal\ (2001) are applied.  This
provides one piece of evidence that the changes made to the Key
Project distance scale are, in fact, improvements.  A stronger
test of the HST Cepheid distance scale based on the maser distance to
NGC~4258 would require a substantially larger sample of Cepheids
(which should reduce the uncertainties in determining the $VI$ P--L
relations and the reddening) and better determination of aperture
corrections; these issues can be addressed simultaneously by searching
for Cepheids with HST in a field that contains more stars and has
undergone more recent star formation, preferably with the higher
resolution that will be afforded by the Advanced Camera for Surveys.
It would be reasonable to expect that observations of a region richer
in Cepheids might yield as many as 3 times the number of Cepheids
(giving a distance modulus uncertainty of 0.04 magnitude) and aperture
corrections accurate to $\pm 0.04$ magnitude; better agreement between
ALLFRAME and DoPHOT analyses might also occur with improved
data. Reductions of uncertainties in the maser distance (e.g., via
improved constraints on the eccentricity of the circumnuclear disk),
would also be greatly beneficial for its use to calibrate the
extragalactic distance scale.  Successful maser distances to other
galaxies, establishing a Hubble relation, would more firmly establish
this novel and promising technique.  The data available at present are
sufficient only to test calibrations of the extragalactic distance
scale, but that alone is of great value.  With improvements in both
Cepheid and maser analyses, NGC~4258 has great potential for
establishing a new primary step in the distance ladder, reducing the
potential systematic errors in measurements of the Hubble constant to
perhaps as little as 5\%, and bypassing controversies over the
distance to the Large Magellanic cloud and the effect of metallicity
on the colors and magnitudes of Cepheids entriely.

\acknowledgements{ We would like to thank Bryan Mendez for his
assistance in attempting a TRGB measurement of the distance to NGC
4258 and our referee and editor for their helpful suggestions.  This
work was supported by NASA grant GO-07227 from the Space Telescope
Science Institute (operated by AURA, Inc. under NASA contract NAS
5-26555). }

\clearpage
\pagestyle{empty}

\clearpage

\clearpage

\figcaption{A $13 \arcmin \times 13 \arcmin$ Digital Sky Survey image of NGC~4258, with the field observed using HST superimposed.  North is oriented vertically and East to the left in this image.}
\plotone{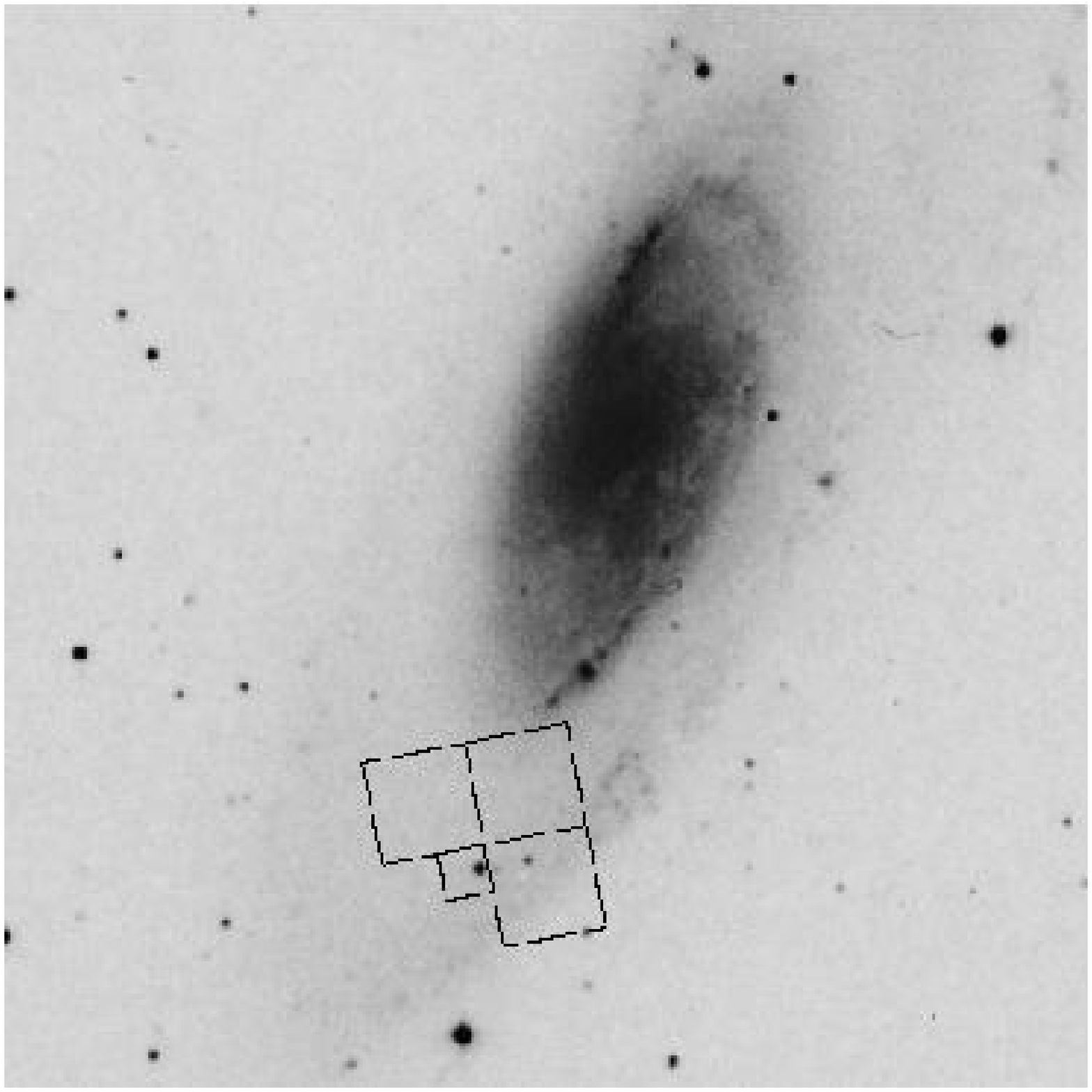}

\clearpage
\figcaption{a) An image of the field covered by WF chip 2 obtained by coadding all images.  The candidate Cepheids found on WF2 are circled and labelled.  b) As a), but WF3 is shown. }

\plotone{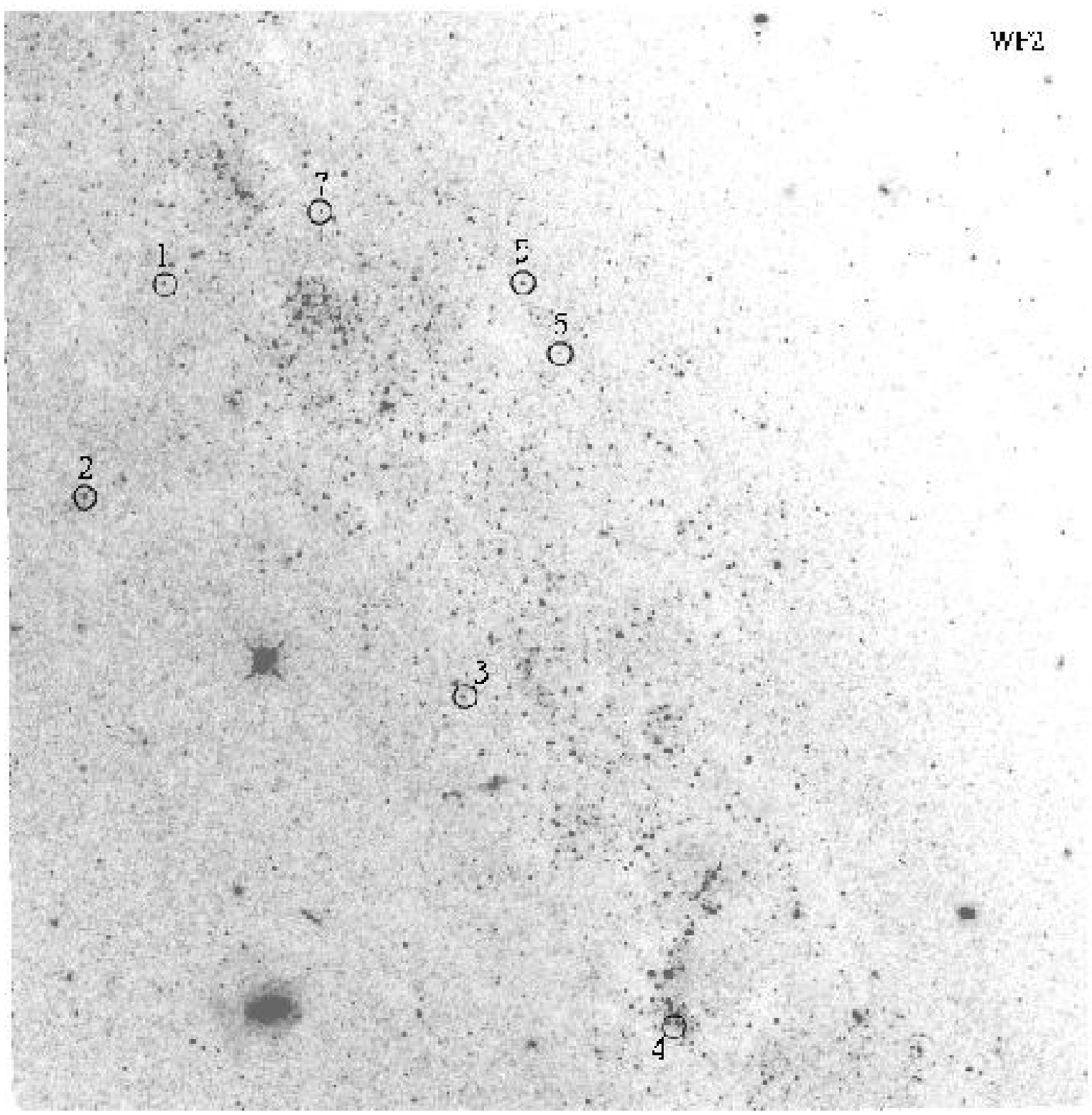}

\clearpage
\plotone{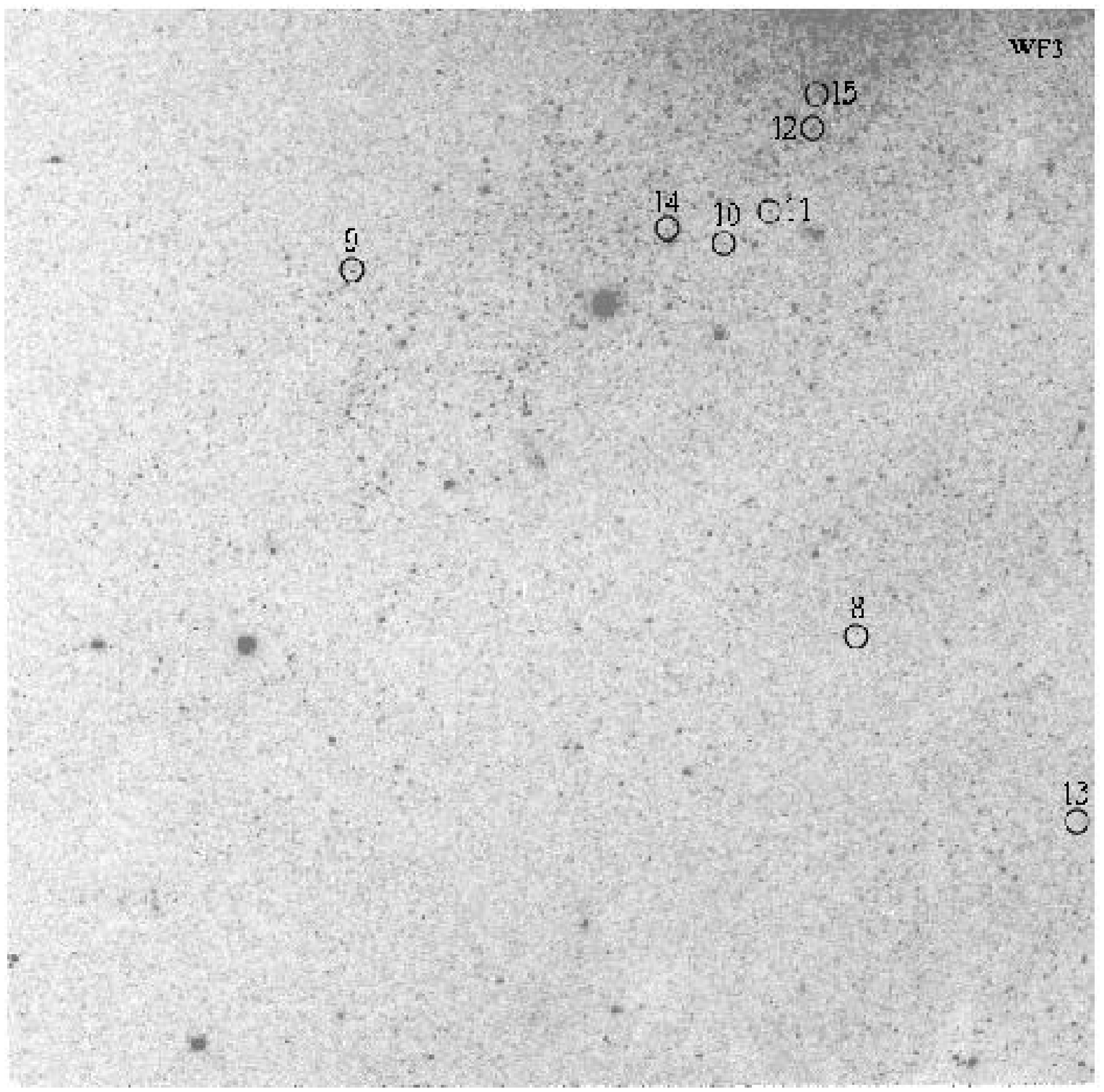}

\clearpage
\figcaption{Finding charts for the candidate Cepheids found, labelled as in Fig. 2.  Each subimage shown is $7 \arcsec \times 7 \arcsec$, and oriented as the images in Fig. 2.}
\epsscale{0.85}
\plotone{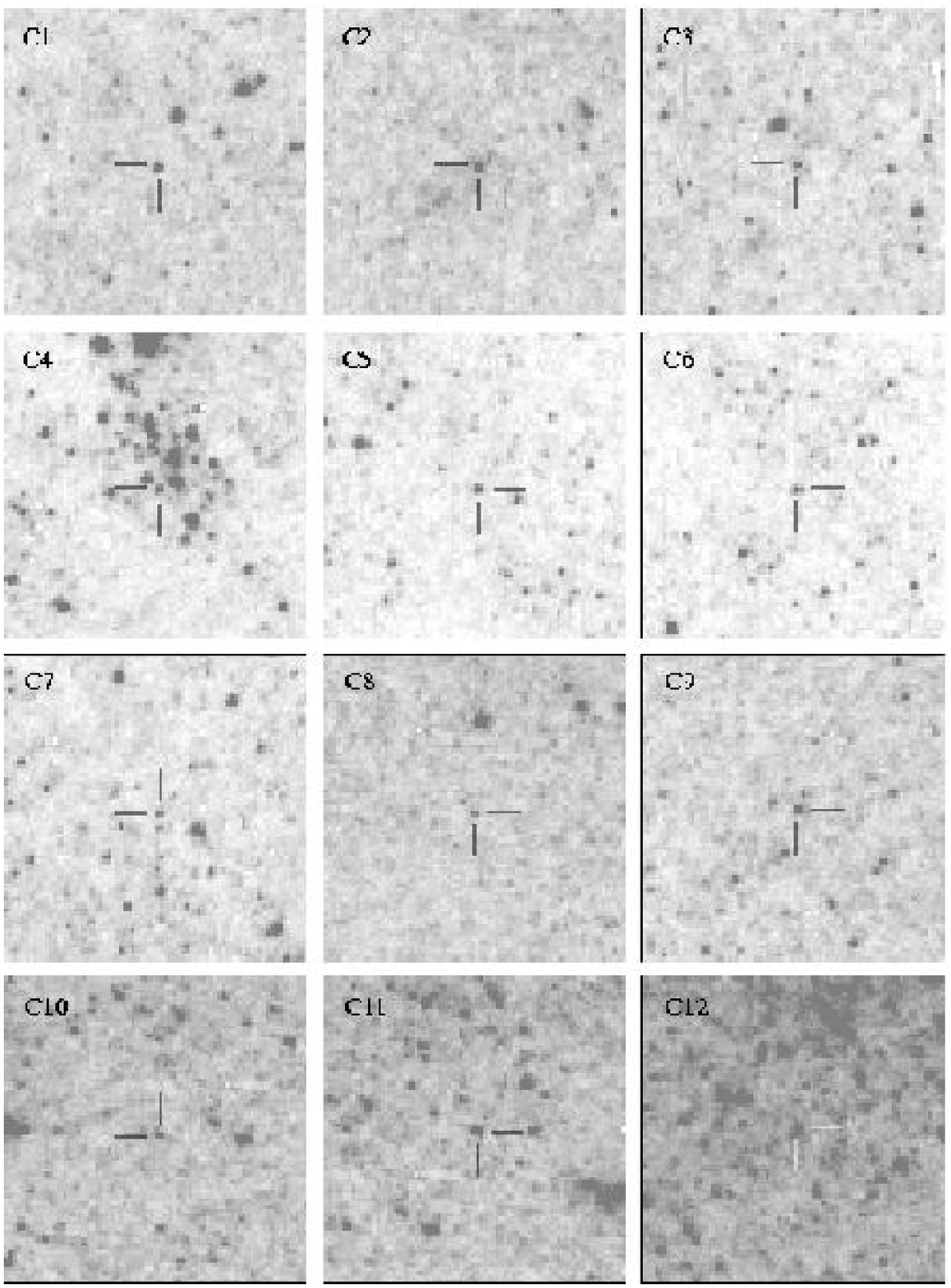}
\clearpage
\plotone{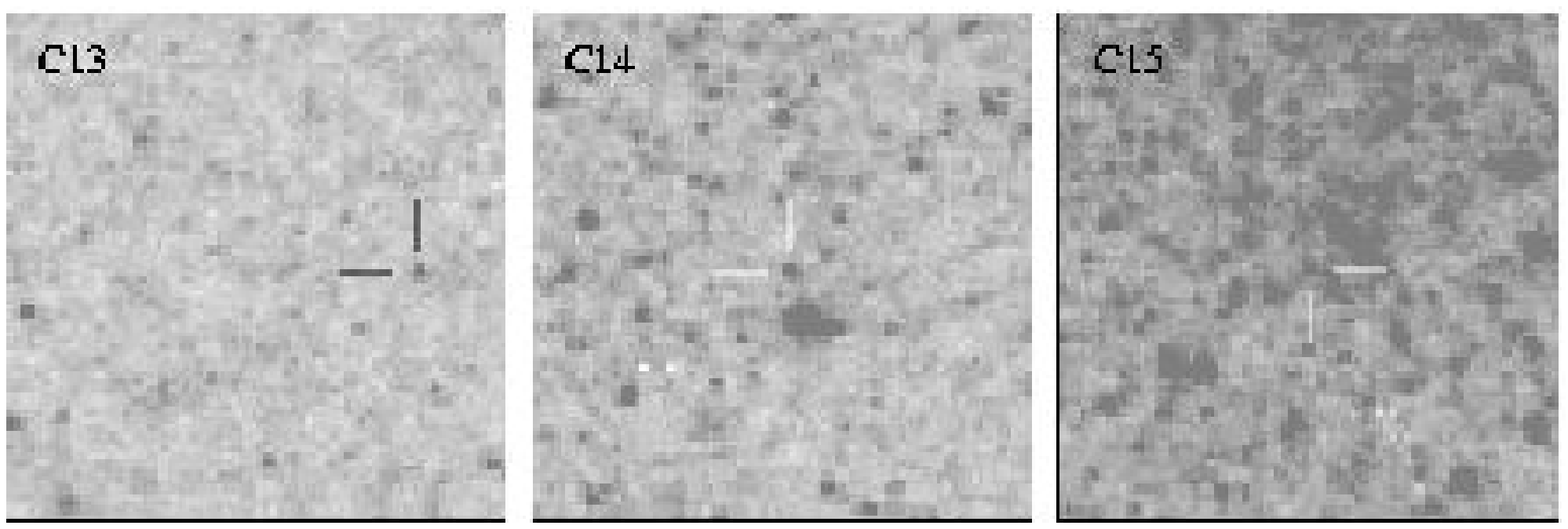}

\epsscale{1}
\clearpage
\figcaption{DoPHOT $F555W$ (left) and $F814W$ (right) light curves for the candidate Cepheids found, plotted versus phase of variation.  The period (in days) for each Cepheid determined during the DoPHOT analysis is also listed in its label.}

\epsscale{0.75}
\clearpage
\figcaption{(top) DoPHOT $V$-band Period--Luminosity relation for NGC~4258.  Cepheids on WF 2 are denoted with an open diamond, those on WF 3 with a filled circle. The solid line is the best-fit P--L relation with slope as in Freedman \etal\ (2001).  (bottom) As above, but based on DoPHOT $I$ data.}
\plotone{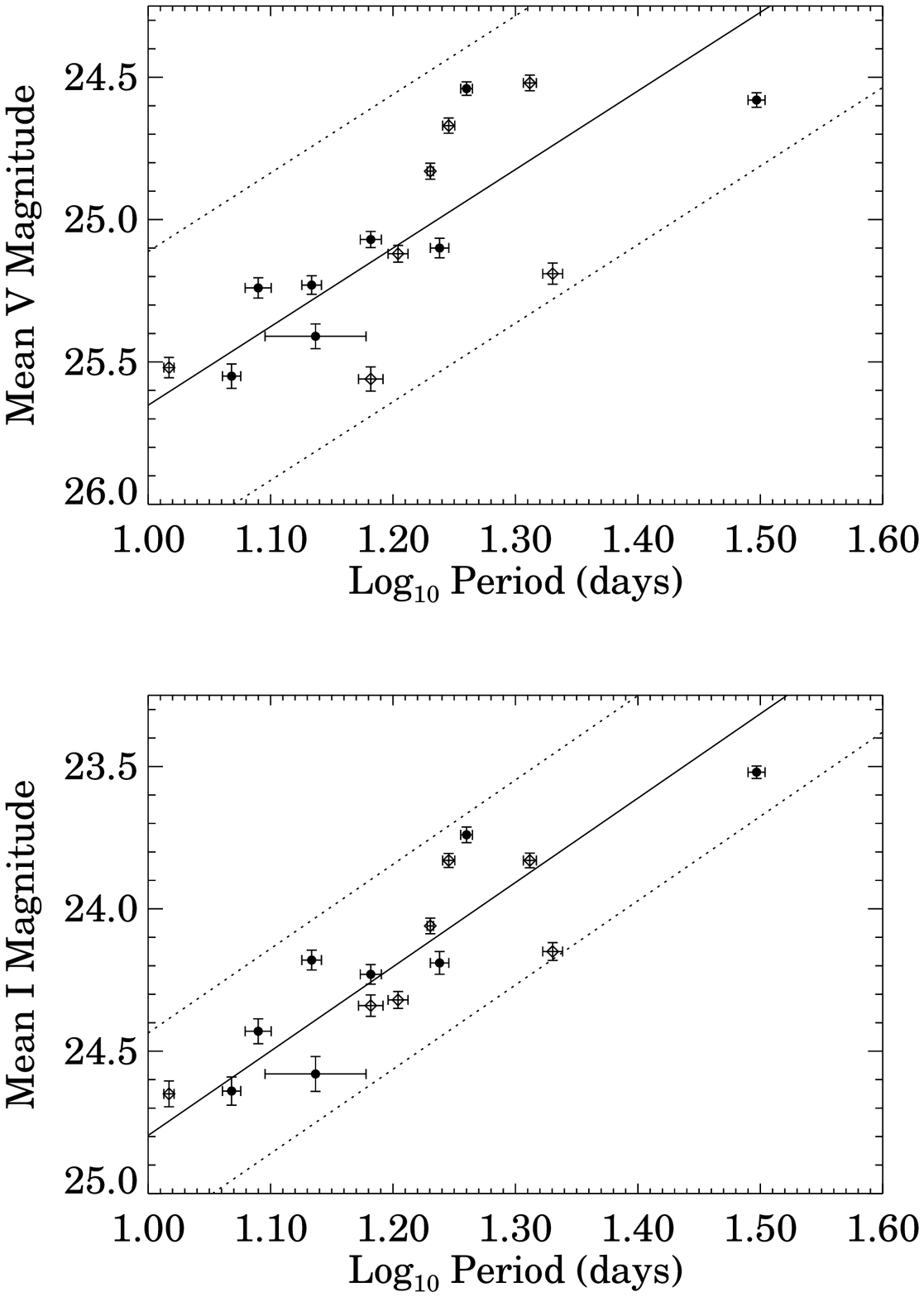}

\clearpage
\figcaption{(top) ALLFRAME $V$-band Period--Luminosity relation for NGC~4258.  Cepheids on WF 2 are denoted with an open diamond, those on WF 3 with a filled circle. The solid line is the best-fit P--L relation with slope as in Freedman \etal\ (2001).  (bottom) As above, but based on ALLFRAME $I$ data.  The WF3/$I$ discrepancy of $\sim 0.06$ mag cannot readily be seen on this plot.} 
\plotone{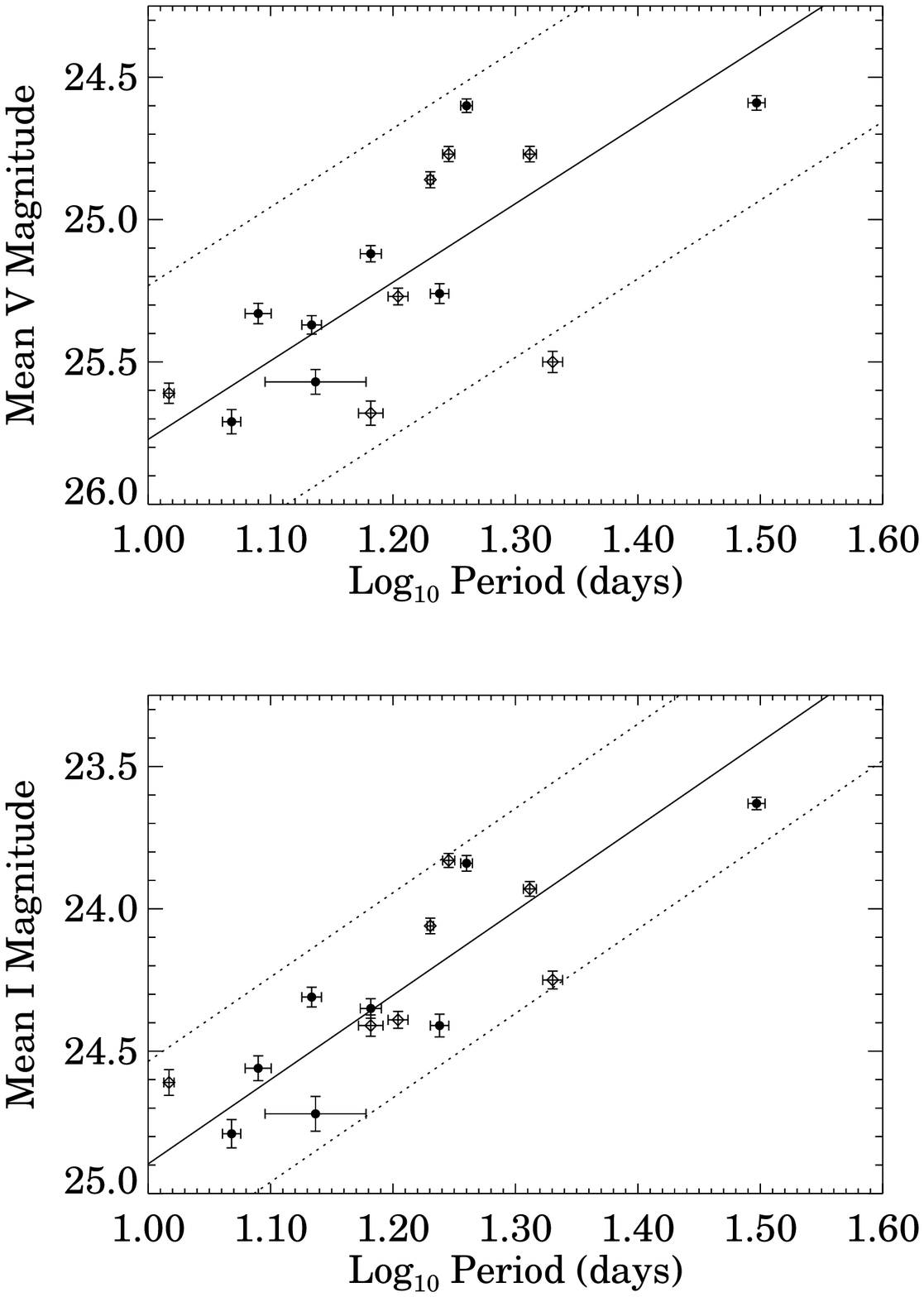}

\clearpage
\begin{deluxetable}{lccccc}
\tablecolumns{6}
\tablewidth{0pc}
\tabletypesize{\small}
\tablenum{1}
\tablecaption{Journal of Observations\tablenotemark{1} \label{tbl-1}}
\tablehead{
\colhead{Epoch} &
\colhead{Filter Used} &
\colhead{Date} &
\colhead{Julian Date\tablenotemark{2}} &
\colhead{Exposure Time (s)\tablenotemark{3}} &
\colhead{Filenames} 
}
\startdata
      1 &   F555W &      1998 Apr 20 &     4901847.600 & 1000 &     U4F60101R/2R \\ 
        &   F814W &                  &     4901847.632 & 1000 &     U4F60103R/4R \\ 
      2 &   F555W &      1998 Apr 21 &     4901850.428 & 1000 &     U4F60201R/2R \\ 
        &   F814W &                  &     4901850.460 & 1000 &     U4F60203R/4R \\ 
      3 &   F555W &      1998 Apr 23 &     4901854.189 & 1000 &     U4F60301R/2R \\ 
        &   F814W &                  &     4901854.221 & 1000 &     U4F60303R/4R \\ 
      4 &   F555W &      1998 Apr 25 &     4901859.086 & 1000 &     U4F60401R/2R \\ 
        &   F814W &                  &     4901858.118 & 1000 &     U4F60403R/4R \\ 
      5 &   F555W &      1998 Apr 28 &     4901863.866 & 1000 &     U4F60501R/2R \\ 
        &   F814W &                  &     4901863.898 & 1000 &     U4F60503R/4R \\ 
      6 &   F555W &       1998 May 1 &     4901869.241 & 1000 &     U4F60601R/2R \\ 
        &   F814W &                  &     4901869.273 & 1000 &     U4F60603R/4R \\ 
      7 &   F555W &       1998 May 5 &     4901878.034 & 1000 &     U4F60701R/2R \\ 
        &   F814W &                  &     4901878.066 & 1000 &     U4F60703R/4R \\ 
      8 &   F555W &      1998 May 10 &     4901887.113 & 1000 &     U4F60801R/2R \\ 
        &   F814W &                  &     4901887.145 & 1000 &     U4F60803R/4R \\ 
      9 &   F555W &      1998 May 15 &     4901897.864 & 1000 &     U4F60901R/2R \\ 
        &   F814W &                  &     4901897.896 & 1000 &     U4F60903R/4R \\ 
     10 &   F555W &      1998 May 22 &     4901912.376 & 1000 &     U4F61001R/2R \\ 
        &   F814W &                  &     4901912.408 & 1000 &     U4F61003R/4R \\ 
     11 &   F555W &      1998 May 30 &     4901928.500 & 1000 &     U4F61101R/2R \\ 
        &   F814W &                  &     4901928.532 & 1000 &     U4F61103R/4R \\ 

\enddata
\tablenotetext{1}{For all observations, the WFPC2 camera was used centered on R.A. 12$^h$ 19$^m$ 8.95$^s$, declination 47$^d$ 13$^m$ 26.76$^s$ (J2000), with roll angle 325.9943} 
\tablenotetext{2}{Heliocentric; at midpoint of the two exposures} 
\tablenotetext{3}{Total of two exposures} 
\end{deluxetable}

\clearpage
\begin{deluxetable}{lccc|cccc|ccc}
\setlength{\tabcolsep}{0.06in} 
\tabletypesize{\small}
\tablecolumns{11}
\tablewidth{0pc}
\tablenum{2}
\tablecaption{Parameters of Cepheids Found \label{tbl-2}}
\tablehead{
\multicolumn{4}{c}{} & \multicolumn{4}{c}{DoPHOT} & \multicolumn{3}{c}{ALLFRAME} \\
\colhead{ID} &
\colhead{Chip} &
\colhead{X\tablenotemark{1}} &
\colhead{Y\tablenotemark{1}} &
\colhead{$<V>$\tablenotemark{2}} &
\colhead{$<I>$\tablenotemark{2}} &
\colhead{Period (d)} &
\colhead{Amplitude\tablenotemark{3}} &
\colhead{$<V>$\tablenotemark{2}} &
\colhead{$<I>$\tablenotemark{2}} &
\colhead{Period (d)}
}
\startdata
C1 & 2 &  158.18 &  603.73 & 24.55 $\pm$ 0.03 & 23.84 $\pm$ 0.03 & 
 20.5 $\pm$ 0.2 & 1.08 & 24.77 & 23.94 &  21.3 \\
C2 & 2 &  102.72 &  455.97 & 24.70 $\pm$ 0.03 & 23.84 $\pm$ 0.02 & 
 17.6 $\pm$ 0.2 & 0.99 & 24.77 & 23.84 &  17.2 \\
C3 & 2 &  364.28 &  317.39 & 24.86 $\pm$ 0.03 & 24.07 $\pm$ 0.03 & 
 17.0 $\pm$ 0.1 & 1.14 & 24.86 & 24.07 &  17.0 \\
C4 & 2 &  509.26 &   90.23 & 25.22 $\pm$ 0.04 & 24.16 $\pm$ 0.03 & 
 21.4 $\pm$ 0.4 & 1.00 & 25.50 & 24.26 &  20.5 \\
C5 & 2 &  405.30 &  604.14 & 25.15 $\pm$ 0.03 & 24.33 $\pm$ 0.03 & 
 16.0 $\pm$ 0.3 & 0.68 & 25.27 & 24.40 &  14.0 \\
C6 & 2 &  431.21 &  554.94 & 25.59 $\pm$ 0.04 & 24.35 $\pm$ 0.04 & 
 15.2 $\pm$ 0.3 & 0.74 & 25.68 & 24.42 &  13.9 \\
C7 & 2 &  447.29 &  566.43 & 25.55 $\pm$ 0.04 & 24.66 $\pm$ 0.05 & 
 10.4 $\pm$ 0.1 & 0.79 & 25.61 & 24.62 &  10.1 \\
C8 & 3 &  627.67 &  365.75 & 25.58 $\pm$ 0.04 & 24.65 $\pm$ 0.05 & 
 11.7 $\pm$ 0.2 & 0.60 & 25.71 & 24.80 &  11.2 \\
C9 & 3 &  276.75 &  618.19 & 25.10 $\pm$ 0.03 & 24.24 $\pm$ 0.03 & 
 15.2 $\pm$ 0.3 & 0.95 & 25.12 & 24.36 &  15.2 \\
C10 & 3 &  535.47 &  636.44 & 25.13 $\pm$ 0.03 & 24.20 $\pm$ 0.04 & 
 17.3 $\pm$ 0.3 & 0.81 & 25.26 & 24.42 &  17.6 \\
C11 & 3 &  566.89 &  659.25 & 24.61 $\pm$ 0.03 & 23.53 $\pm$ 0.02 & 
 31.4 $\pm$ 0.5 & 0.82 & 24.59 & 23.64 &  30.7 \\
C12 & 3 &  597.51 &  717.16 & 24.57 $\pm$ 0.02 & 23.75 $\pm$ 0.03 & 
 18.2 $\pm$ 0.2 & 0.93 & 24.60 & 23.85 &  17.6 \\
C13 & 3 &  780.74 &  236.12 & 25.27 $\pm$ 0.04 & 24.44 $\pm$ 0.04 & 
 12.3 $\pm$ 0.3 & 0.97 & 25.33 & 24.57 &  12.4 \\
C14 & 3 &  495.77 &  645.99 & 25.26 $\pm$ 0.03 & 24.19 $\pm$ 0.03 & 
 13.6 $\pm$ 0.2 & 0.74 & 25.37 & 24.32 &  13.7 \\
C15 & 3 &  599.59 &  740.72 & 25.44 $\pm$ 0.04 & 24.59 $\pm$ 0.06 & 
 13.7 $\pm$ 1.3 & 0.82 & 25.57 & 24.73 &  13.0 \\
\enddata
\tablenotetext{1}{In the coordinate system of the first $F555W$ exposure, U4F60101R} 
\tablenotetext{2}{Intensity-weighted mean magnitudes are given for DoPHOT; for ALLFRAME, intensity-weighted means based on the template fits are listed}
\tablenotetext{3}{Peak-to-peak, in magnitudes, based upon template fit}
\end{deluxetable}

\clearpage
\begin{deluxetable}{lccccc}
\tablecolumns{6}
\tablewidth{0pc}
\footnotesize
\tablenum{3}
\tablecaption{DoPHOT Photometry of Cepheids Found \label{tbl-3a}}
\tablehead{
\colhead{Epoch} &
\colhead{C1} &
\colhead{C2} &
\colhead{C3} &
\colhead{C4} &
\colhead{C5} 
}
\startdata

\multicolumn{6}{c}{$V$} \\
1 & 25.07$\pm$0.10 & 25.26$\pm$0.11 & 25.12$\pm$0.10 & 24.97$\pm$0.07 & 
25.27$\pm$0.09 \\
2 & 24.11$\pm$0.05 & 24.99$\pm$0.10 & 25.27$\pm$0.12 & 24.99$\pm$0.09 & 
24.80$\pm$0.08 \\
3 & 24.06$\pm$0.04 & 24.10$\pm$0.08 & 25.41$\pm$0.11 & 25.07$\pm$0.10 & 
24.73$\pm$0.06 \\
4 & 24.33$\pm$0.06 & 24.46$\pm$0.08 & 25.30$\pm$0.10 & 25.13$\pm$0.09 & 
25.15$\pm$0.08 \\
5 & 24.52$\pm$0.06 & 24.58$\pm$0.06 & 24.31$\pm$0.06 & 25.56$\pm$0.12 & 
25.27$\pm$0.11 \\
6 & 24.70$\pm$0.08 & 24.82$\pm$0.08 & 24.47$\pm$0.12 & 26.02$\pm$0.18 & 
25.60$\pm$0.12 \\
7 & 25.09$\pm$0.09 & 25.41$\pm$0.11 & 24.72$\pm$0.08 & 25.77$\pm$0.12 & 
25.20$\pm$0.11 \\
8 & \dots & 24.72$\pm$0.09 & 25.40$\pm$0.12 & 24.75$\pm$0.07 & 
\dots \\
9 & 24.34$\pm$0.09 & 24.55$\pm$0.06 & 24.25$\pm$0.08 & 25.44$\pm$0.12 & 
25.40$\pm$0.10 \\
10 & 24.76$\pm$0.08 & 25.22$\pm$0.10 & 24.91$\pm$0.09 & 25.99$\pm$0.15 & 
25.09$\pm$0.08 \\
11 & 25.24$\pm$0.09 & 24.38$\pm$0.06 & 25.20$\pm$0.11 & 24.69$\pm$0.07 & 
25.30$\pm$0.10 \\
\multicolumn{6}{c}{$I$} \\

1 & 24.25$\pm$0.13 & 24.15$\pm$0.08 & 23.95$\pm$0.09 & 23.95$\pm$0.08 & 
24.34$\pm$0.09 \\
2 & 23.75$\pm$0.07 & 23.95$\pm$0.08 & 24.16$\pm$0.10 & 24.05$\pm$0.09 & 
24.34$\pm$0.09 \\
3 & 23.48$\pm$0.08 & 23.63$\pm$0.08 & 24.24$\pm$0.09 & 23.95$\pm$0.08 & 
24.06$\pm$0.10 \\
4 & 23.74$\pm$0.09 & 23.50$\pm$0.10 & 24.38$\pm$0.12 & 24.10$\pm$0.09 & 
24.32$\pm$0.10 \\
5 & 23.73$\pm$0.08 & 23.70$\pm$0.08 & 23.80$\pm$0.07 & 24.27$\pm$0.08 & 
24.39$\pm$0.11 \\
6 & 23.83$\pm$0.09 & 23.87$\pm$0.08 & 24.06$\pm$0.12 & 24.61$\pm$0.13 & 
24.49$\pm$0.10 \\
7 & 23.97$\pm$0.09 & 24.07$\pm$0.08 & 23.96$\pm$0.09 & 24.77$\pm$0.12 & 
24.40$\pm$0.10 \\
8 & 24.20$\pm$0.09 & 23.82$\pm$0.07 & 24.42$\pm$0.12 & 23.92$\pm$0.07 & 
24.05$\pm$0.08 \\
9 & 23.52$\pm$0.06 & 23.79$\pm$0.08 & 23.67$\pm$0.07 & 24.07$\pm$0.08 & 
24.75$\pm$0.12 \\
10 & 23.90$\pm$0.07 & 24.14$\pm$0.09 & 23.94$\pm$0.08 & 24.65$\pm$0.11 & 
24.23$\pm$0.08 \\
11 & 24.09$\pm$0.08 & 23.48$\pm$0.08 & 24.50$\pm$0.12 & 23.91$\pm$0.06 & 
24.39$\pm$0.13 \\

\enddata
\end{deluxetable}

\clearpage
\begin{deluxetable}{lccccc}
\tablecolumns{6}
\tablewidth{0pc}
\footnotesize
\tablenum{3}
\tablecaption{DoPHOT Photometry of Cepheids Found \label{tbl-3b}}
\tablehead{
\colhead{Epoch} &
\colhead{C6} &
\colhead{C7} &
\colhead{C8} &
\colhead{C9} &
\colhead{C10} 
}
\startdata
\multicolumn{6}{c}{$V$} \\
1 & 25.32$\pm$0.10 & 25.56$\pm$0.14 & 25.57$\pm$0.13 & 24.77$\pm$0.08 & 
24.64$\pm$0.06 \\
2 & 25.28$\pm$0.11 & 25.87$\pm$0.15 & 25.51$\pm$0.11 & 24.84$\pm$0.08 & 
24.78$\pm$0.08 \\
3 & 25.35$\pm$0.11 & 25.62$\pm$0.11 & 25.95$\pm$0.17 & 25.07$\pm$0.08 & 
25.02$\pm$0.08 \\
4 & 25.79$\pm$0.15 & 25.23$\pm$0.09 & 25.77$\pm$0.17 & 25.34$\pm$0.11 & 
25.07$\pm$0.09 \\
5 & 26.07$\pm$0.18 & 25.25$\pm$0.10 & 25.24$\pm$0.10 & 25.60$\pm$0.15 & 
25.47$\pm$0.15 \\
6 & 25.91$\pm$0.13 & 25.94$\pm$0.17 & 25.37$\pm$0.11 & 25.16$\pm$0.10 & 
25.43$\pm$0.12 \\
7 & 25.30$\pm$0.09 & 25.60$\pm$0.11 & 25.98$\pm$0.19 & 24.63$\pm$0.06 & 
25.41$\pm$0.13 \\
8 & 25.67$\pm$0.13 & 25.55$\pm$0.12 & 25.33$\pm$0.10 & 25.31$\pm$0.09 & 
24.92$\pm$0.12 \\
9 & 25.88$\pm$0.16 & 25.46$\pm$0.10 & 25.78$\pm$0.14 & 25.33$\pm$0.11 & 
25.37$\pm$0.10 \\
10 & 25.30$\pm$0.13 & 25.86$\pm$0.17 & 25.37$\pm$0.12 & 24.96$\pm$0.09 & 
25.50$\pm$0.12 \\
11 & 26.06$\pm$0.16 & 25.37$\pm$0.12 & 25.85$\pm$0.18 & 25.60$\pm$0.14 & 
25.18$\pm$0.14 \\
\multicolumn{6}{c}{$I$} \\

1 & 24.41$\pm$0.11 & 24.46$\pm$0.14 & 24.51$\pm$0.14 & 24.20$\pm$0.11 & 
24.05$\pm$0.10 \\
2 & 24.28$\pm$0.12 & 24.84$\pm$0.18 & 24.69$\pm$0.17 & 23.81$\pm$0.09 & 
23.90$\pm$0.08 \\
3 & 24.20$\pm$0.13 & 24.89$\pm$0.15 & 24.82$\pm$0.18 & 24.18$\pm$0.10 & 
24.04$\pm$0.10 \\
4 & 24.36$\pm$0.13 & 24.53$\pm$0.13 & 24.61$\pm$0.15 & 24.24$\pm$0.12 & 
24.07$\pm$0.10 \\
5 & 24.24$\pm$0.12 & 24.55$\pm$0.11 & 24.60$\pm$0.15 & 24.70$\pm$0.18 & 
24.30$\pm$0.12 \\
6 & 24.63$\pm$0.15 & 24.82$\pm$0.18 & 24.42$\pm$0.12 & 24.49$\pm$0.13 & 
24.55$\pm$0.14 \\
7 & 24.14$\pm$0.10 & 24.55$\pm$0.12 & 25.01$\pm$0.23 & 24.13$\pm$0.11 & 
24.46$\pm$0.14 \\
8 & 24.20$\pm$0.11 & 24.78$\pm$0.16 & 24.31$\pm$0.13 & 24.13$\pm$0.09 & 
23.94$\pm$0.11 \\
9 & 24.81$\pm$0.15 & 24.46$\pm$0.14 & 25.13$\pm$0.22 & 24.59$\pm$0.13 & 
24.26$\pm$0.11 \\
10 & 24.39$\pm$0.12 & 24.73$\pm$0.20 & 24.59$\pm$0.17 & 24.15$\pm$0.10 & 
24.61$\pm$0.17 \\
11 & 24.36$\pm$0.11 & 24.78$\pm$0.21 & 24.73$\pm$0.16 & 24.37$\pm$0.13 & 
24.29$\pm$0.14 \\

\enddata
\end{deluxetable}

\clearpage
\begin{deluxetable}{lccccc}
\tablecolumns{6}
\tablewidth{0pc}
\footnotesize
\tablenum{3}
\tablecaption{DoPHOT Photometry of Cepheids Found \label{tbl-3c}}
\tablehead{
\colhead{Epoch} &
\colhead{C11} &
\colhead{C12} &
\colhead{C13} &
\colhead{C14} &
\colhead{C15} 
}
\startdata
\multicolumn{6}{c}{$V$} \\
1 & 24.83$\pm$0.14 & 24.89$\pm$0.11 & 24.82$\pm$0.10 & 25.50$\pm$0.16 & 
25.16$\pm$0.11 \\
2 & 24.86$\pm$0.08 & 24.87$\pm$0.08 & 25.10$\pm$0.09 & 25.11$\pm$0.09 & 
25.16$\pm$0.10 \\
3 & 24.95$\pm$0.09 & 24.95$\pm$0.08 & 25.32$\pm$0.12 & 24.76$\pm$0.09 & 
25.29$\pm$0.11 \\
4 & 24.87$\pm$0.09 & 24.95$\pm$0.08 & 25.80$\pm$0.14 & 25.00$\pm$0.08 & 
25.69$\pm$0.17 \\
5 & 25.01$\pm$0.10 & 24.04$\pm$0.06 & 25.52$\pm$0.12 & 25.27$\pm$0.12 & 
26.04$\pm$0.24 \\
6 & 24.47$\pm$0.08 & 24.37$\pm$0.09 & 24.80$\pm$0.08 & 25.43$\pm$0.12 & 
25.49$\pm$0.14 \\
7 & 24.02$\pm$0.08 & 24.57$\pm$0.08 & 25.35$\pm$0.10 & 25.52$\pm$0.10 & 
25.24$\pm$0.11 \\
8 & 24.17$\pm$0.08 & 24.94$\pm$0.09 & 25.77$\pm$0.15 & 25.32$\pm$0.10 & 
25.90$\pm$0.17 \\
9 & 24.54$\pm$0.06 & 23.89$\pm$0.05 & 24.97$\pm$0.08 & 25.70$\pm$0.14 & 
25.37$\pm$0.13 \\
10 & 24.78$\pm$0.07 & 24.63$\pm$0.08 & 25.88$\pm$0.16 & 25.02$\pm$0.08 & 
25.79$\pm$0.15 \\
11 & 24.72$\pm$0.08 & 24.88$\pm$0.11 & 25.33$\pm$0.12 & 25.68$\pm$0.12 & 
25.17$\pm$0.11 \\

\multicolumn{6}{c}{$I$} \\
1 & 23.71$\pm$0.09 & 24.03$\pm$0.11 & 24.47$\pm$0.17 & 24.30$\pm$0.12 & 
24.25$\pm$0.14 \\
2 & 23.53$\pm$0.09 & 23.96$\pm$0.10 & 24.33$\pm$0.13 & 24.16$\pm$0.10 & 
24.54$\pm$0.16 \\
3 & 23.79$\pm$0.07 & 24.10$\pm$0.12 & 24.64$\pm$0.16 & 23.95$\pm$0.10 & 
24.35$\pm$0.13 \\
4 & 23.74$\pm$0.07 & 23.57$\pm$0.09 & 24.72$\pm$0.15 & 24.00$\pm$0.12 & 
24.75$\pm$0.22 \\
5 & 23.77$\pm$0.07 & 23.36$\pm$0.07 & 24.56$\pm$0.14 & 24.36$\pm$0.14 & 
25.03$\pm$0.26 \\
6 & 23.44$\pm$0.06 & 23.58$\pm$0.08 & 24.16$\pm$0.11 & 24.24$\pm$0.12 & 
24.98$\pm$0.25 \\
7 & 23.33$\pm$0.09 & 23.77$\pm$0.08 & 24.12$\pm$0.13 & 24.18$\pm$0.11 & 
24.52$\pm$0.17 \\
8 & 23.22$\pm$0.05 & 23.97$\pm$0.11 & 24.77$\pm$0.17 & 24.12$\pm$0.11 & 
\dots \\
9 & 23.37$\pm$0.06 & 23.39$\pm$0.07 & 24.46$\pm$0.16 & 24.49$\pm$0.14 & 
24.41$\pm$0.13 \\
10 & 23.59$\pm$0.07 & 23.94$\pm$0.12 & 24.54$\pm$0.17 & 24.18$\pm$0.10 & 
\dots \\
11 & 23.60$\pm$0.06 & 23.86$\pm$0.10 & 24.29$\pm$0.12 & 24.23$\pm$0.13 & 
24.43$\pm$0.17 \\

\enddata
\end{deluxetable}

\clearpage
\begin{deluxetable}{lclll}
\tablecolumns{5}
\tablewidth{0pc}
\tabletypesize{\normalsize}
\tablenum{4}
\tablecaption{NGC~4258 Distance Moduli (no metallicity correction) \label{tbl-4}}
\tablehead{
\colhead{Photometry Used} &
\colhead{Subset of Cepheids} &
\colhead{$\mu_V$ (mag)} &
\colhead{$\mu_I$ (mag)} &
\colhead{$\mu_0$ (mag)} 
}
\startdata
{\bf DoPHOT} & All & 29.90 $\pm$ 0.07 & 29.69 $\pm$ 0.05& {\bf 29.40 $\pm$ 0.06 }\\
 & Chip 2 & 29.92 $\pm$ 0.13 & 29.72 $\pm$ 0.085 & 29.43 $\pm$ 0.10 \\
 & Chip 3 & 29.87 $\pm$ 0.10 & 29.67 $\pm$ 0.06 & 29.38 $\pm$ 0.07 \\
{\bf ALLFRAME} & All & 29.99 $\pm$ 0.075 & 29.80 $\pm$ 0.05 & {\bf 29.53 $\pm$ 0.07} \\
 & Chip 2 & 30.03 $\pm$ 0.14 & 29.77 $\pm$ 0.09 & 29.40 $\pm$ 0.10 \\
 & Chip 3 & 29.97 $\pm$ 0.08 & 29.84 $\pm$ 0.07 & 29.65 $\pm$ 0.09 \\
{\bf Corrected ALLFRAME} & All & 29.99 $\pm$ 0.075 & 29.77 $\pm$ 0.05 & {\bf 29.44 $\pm$ 0.06} \\
 & Chip 2 & 30.03 $\pm$ 0.14 & 29.77 $\pm$ 0.09 & 29.40 $\pm$ 0.10 \\
 & Chip 3 & 29.97 $\pm$ 0.08 & 29.77 $\pm$ 0.07 & 29.48 $\pm$ 0.09 \\
\enddata
\end{deluxetable}

\clearpage
\begin{deluxetable}{lll}
\tablecolumns{3}
\tablewidth{0pc}
\tighten
\tablenum{5}
\tabletypesize{\normalsize}
\tablecaption{{Uncertainties in This Distance Measurement}}
\tablehead{
\colhead{ } &
\colhead{Source} &
\colhead{Error} 
}
\startdata
S$_{1}$ & Systematic Errors in LMC P-L Calibration &  \\
   &{\it A. LMC True Modulus} &  $\pm$0.10  \\ 
    &{\it B. LMC P-L Zero Point} & $\pm$0.02  \\
   & \hspace{.2in} A and B added in quadrature & $\pm$0.10 \\
\\[1pt]
S$_{2}$ & Systematic Errors in WFPC2 Zero Points & $\pm 0.07 $ \\
\\[1pt]
S$_{3}$ & Average Metallicity Correction & $\pm$0.08 \\
\\[4pt]
S$_{4}$ & Systematic Errors Unique to NGC~4258 Photometry & $\pm$0.05  \\
\\[4pt]
S$_{5}$ & Dependence of NGC~4258 Distance Modulus on Magnitude Averaging Method & $\pm$0.04  \\
\\[4pt]
R$_{1}$ & Random Error in the NGC~4258 Extinction-Corrected Distance Modulus & \\
   &{\it C. NGC~4258 P-L Fit (V)} &  $\pm$0.07   \\
   &{\it D. NGC~4258 P-L Fit (I)} &  $\pm$0.05   \\
   & \hspace{.2in} C and D partially correlated & $\pm$0.06 \\
\\[6pt]
R$_{tot}$  & Errors Only Affecting This Determination & {\bf $\pm$0.09} \\
 & \hspace{.2in}(S$_{4}$, S$_{5}$, and R$_{1}$ added in quadrature) & \\
S$_{tot}$  & Systematic Errors in Key Project Techniques &{\bf $\pm$0.15} \\
 & \hspace{.2in}(S$_{1}$, S$_{2}$, and S$_{3}$ added in quadrature) & \\

\enddata


\end{deluxetable}


\end{document}